\documentclass[aps,preprint,showpacs,amsmath,amssymb,amsfonts]{revtex4}
\usepackage{epsfig}
\usepackage{graphicx}
\usepackage{subfigure}
\usepackage{dcolumn}
\usepackage{bm}
\usepackage{amsthm}
\usepackage{enumitem}
\usepackage{fixltx2e}
\usepackage{slashed}

\newcommand{\del}{\partial}

\newcommand{\ddsimple}[2]{\frac{\del #1}{\del #2}} 
\renewcommand{\Re}{\operatorname{Re}}
\renewcommand{\Im}{\operatorname{Im}}

\newcommand{\bbR}{\mathbb{R}}
\newcommand{\bbC}{\mathbb{C}}

\newcommand{\bbZ}{\mathbb{Z}}
\newcommand{\calB}{\mathcal{B}}
\newcommand{\calE}{\mathcal{E}}
\newcommand{\calD}{\mathcal{D}}
\newcommand{\scri}{\mathcal{I}}
\newcommand{\scrid}{\scri^{\text{id}}}

\newcommand{\dSCFT}[2]{dS\textsubscript{#1}/CFT\textsubscript{#2}}

\newtheorem{theorem}{Theorem}

\begin{document}

\title{Antipodally symmetric gauge fields and higher-spin gravity\\ in de Sitter space}

\author{Yasha Neiman}
\email{yashula@gmail.com}

\affiliation{Perimeter Institute for Theoretical Physics, 31 Caroline Street N, Waterloo, ON, N2L 2Y5, Canada}

\date{\today}

\begin{abstract}
We study gauge fields of arbitrary spin in de Sitter space. These include Yang-Mills fields and gravitons, as well as the higher-spin fields of Vasiliev theory. We focus on antipodally symmetric solutions to the field equations, i.e. ones that live on ``elliptic'' de Sitter space $dS_4/\bbZ_2$. For free fields, we find spanning sets of such solutions, including boundary-to-bulk propagators. We find that free solutions on $dS_4/\bbZ_2$ can only have one of the two types of boundary data at infinity, meaning that the boundary 2-point functions vanish. In Vasiliev theory, this property persists order by order in the interaction, i.e. the boundary $n$-point functions in $dS_4/\bbZ_2$ all vanish. This implies that a higher-spin dS/CFT based on the Lorentzian $dS_4/\bbZ_2$ action is empty. For more general interacting theories, such as ordinary gravity and Yang-Mills, we can use the free-field result to define a well-posed perturbative initial value problem in $dS_4/\bbZ_2$.
\end{abstract}


\maketitle
\tableofcontents
\newpage

\section{Introduction and summary} \label{sec:intro}

\subsection{Motivation from dS/CFT}

The AdS/CFT correspondence \cite{Aharony:1999ti,Witten:1998qj} offers a non-perturbative model of quantum gravity and a concrete realization of the holographic principle. The correspondence relates a gravitational theory in a (locally, asymptotically) anti-de Sitter space with a conformal quantum field theory (CFT) on its boundary at spatial infinity. A field in the AdS bulk has two possible falloff behaviors at infinity. These are the asymptotic analogs of Neumann and Dirichlet boundary conditions. In the CFT, these two types of boundary data correspond to operators and their conjugate background fields. 

The observed positive value of the cosmological constant implies that de Sitter space $dS_4$ is the more realistic of the maximally symmetric spacetimes. De Sitter space is also an ideal theoretical laboratory for quantum gravity in the presence of causal horizons. Unfortunately, the theoretical understanding of dS is somewhat behind the AdS and flat cases. Conformal infinity in $dS_4$ consists of two spacelike 3-spheres $\scri^\pm$, one in the infinite past and the other in the infinite future. One would like to know more about field asymptotics at $\scri^\pm$ and their physical meaning. A related and more ambitious goal is to formulate dS/CFT - a version of AdS/CFT for positive cosmological constant.

dS/CFT was first considered in \cite{Strominger:2001pn}, with emphasis on the \dSCFT{3}{2} case. A concrete proposal for the physically relevant dimensions \dSCFT{4}{3} was made in \cite{Anninos:2011ui}, by analytically continuing a suitable version of AdS/CFT \cite{Klebanov:2002ja,Sezgin:2003pt,Giombi:2012ms}. The duality proposed in \cite{Anninos:2011ui} relates an $Sp(N)$ vector model on the boundary with higher-spin (Vasiliev) gravity \cite{Vasiliev:1999ba} in the bulk. Vasiliev gravity is an interacting theory with an infinite tower of gauge fields of arbitrarily large spin \cite{Fronsdal:1978rb,Fronsdal:1978vb}, including the spin-2 graviton. The theory is believed to be non-local at cosmological scales. At least perturbatively, it does not reduce to General Relativity in any limit. It is nevertheless worth studying, both as a marvel of mathematical physics and as the only concrete proposal for holography in $dS_4$.

In the early discussions of dS/CFT, some basic questions arose. Since the boundary of de Sitter space is composed of two disjoint pieces $\scri^\pm$, on which manifold does the dual CFT live? What is the bulk interpretation of the CFT correlators? On these issues, the \dSCFT{4}{3} proposal of \cite{Anninos:2011ui} follows the ``Hartle-Hawking-Maldacena'' paradigm developed in \cite{Maldacena:2002vr,Harlow:2011ke}. Only $\scri^+$ plays an explicit role in the duality. The CFT partition function $Z_{\text{CFT}}$ (as a function of background fields) is equated by the duality to a preferred \emph{wavefunction} over the bulk field asymptotics on $\scri^+$. This is the Hartle-Hawking wavefunction, obtained by a path integral over Euclidean modes. Schematically:
\begin{align}
 Z_{\text{CFT}}[\text{sources on }S_3] = \Psi_{\text{HH}}[\text{fields on }\scri^+] \ . \label{eq:HH}
\end{align}
The Euclidean modes used to calculate $\Psi_{\text{HH}}$ can be expressed either as real field configurations on a Euclidean AdS (i.e. hyperbolic) space bounded by $\scri^+$, or as complex configurations on dS with positive frequency in the Bunch-Davies sense.

In \cite{Parikh:2002py,Parikh:2004ux,Parikh:2004wh}, a different kind of dS/CFT was considered. The idea is to identify antipodal points in $dS_4$, yielding the so-called ``elliptic'' de Sitter space $dS_4/\bbZ_2$. Past and future infinity are now identified into a single 3-sphere $\scrid$, which every observer can both see and affect. One can then imagine a CFT on $\scrid$, which calculates ``transition amplitudes'', i.e. Lorentzian path integrals, between a state on $\scri^-$ and the same state on $\scri^+$.

In \cite{Parikh:2002py}, the antipodal identification was motivated by the information puzzles concerning cosmological horizons. In particular, $dS_4/\bbZ_2$ provides a radical realization of horizon complementarity \cite{Susskind:1993if,Dyson:2002pf}: the two sides of each horizon are literally the same. In addition, it was argued in \cite{Parikh:2002py} that in $dS_4/\bbZ_2$, the Hilbert space (as opposed to the transition amplitudes) is observer-dependent, and thus not invariant under the full de Sitter group. This may resolve the puzzle of the finite de Sitter entropy. In this paper, we will not deal with these aspects of $dS_4/\bbZ_2$, but we list them here as further motivation for studying this spacetime.

The present paper's main goal was to explore the idea of dS/CFT for antipodally symmetric transition amplitudes, using the concrete bulk theory from \cite{Anninos:2011ui}, i.e. Vasiliev gravity. Our conclusion is that such a dS/CFT would be empty, at least when the bulk fields are perturbative and classical. Specifically, we find that the boundary $n$-point functions of (Lorentzian) Vasiliev gravity on $dS_4/\bbZ_2$ all vanish. More precisely, the $n$-point functions with boundary conditions that preserve the higher-spin symmetry vanish, while the $n$-point functions with other boundary conditions are ill-defined. This result stems from the fact that interactions in Vasiliev theory can be evaluated at any single point in spacetime \cite{Giombi:2010vg,Colombo:2012jx,Didenko:2012tv,Didenko:2013bj}, while the boundary-to-bulk propagators in $dS_4/\bbZ_2$ are distributions that vanish almost everywhere. 

On our way to the above conclusion, we present results of more general interest on antipodally symmetric gauge fields in $dS_4$, or, equivalently, on gauge fields in $dS_4/\bbZ_2$. We summarize these below, along with the plan of the paper. When discussing spin-$s$ gauge fields, we will often include the scalar with mass $m^2=2$ (in units of the de Sitter radius) as the spin-0 case. Such a field appears in Vasiliev gravity alongside the higher-spin gauge fields.

\subsection{Plan of the paper}

The paper is structured as follows. In section \ref{sec:prelim_bulk}, we review the bulk geometry of real and complex $dS_4$ in the ambient $\bbR^{4,1}$ formalism. We discuss the antipodal map, stressing that it is an operation of the CT type. We review the definitions and field equations for free spin-$s$ gauge fields in $dS_4$. We then review the twistor space of $dS_4$, presented as the spinor space of $\bbR^{4,1}$, and its relation to $SO(3,1)$ spinor fields. This leads to a review of the free gauge field equations in spinor form. In section \ref{sec:prelim_scri}, we review the asymptotic geometry of $dS_4$ and the asymptotic boundary data for spin-$s$ gauge fields.

In section \ref{sec:relations}, we prove our first main result: antipodally symmetric gauge fields in $dS_4$ have just one type of boundary data (electric/magnetic) non-vanishing on $\scri$, depending on the sign of the antipodal symmetry. There is a similar result for $m^2=2$ scalars vis. Dirichlet/Neumann boundary data, which was already noted in \cite{Ng:2012xp} (in addition, the spin-2 result was almost stated in \cite{Anninos:2011jp}). We prove these statements using the free equations' conformal symmetry, along with smoothness through $\scri^\pm$ in the ambient $\bbR^{4,2}$ picture. We then justify the smoothness assumption by presenting a spanning set of solutions that satisfy it explicitly. These solutions are bulk 2-point functions, with the second point on the $EAdS_4$ in the imaginary future/past.

In section \ref{sec:elliptic}, we review the geometry of elliptic de Sitter space $dS_4/\bbZ_2$, and reinterpret the previous section's result in terms of boundary two-point functions in $dS_4/\bbZ_2$. The result then states that the 2-point functions of gauge fields are always vanishing or ill-defined, depending on the choice of boundary conditions. In section \ref{sec:initial_value}, we use the structure of the free equations to formulate a well-posed perturbative initial value problem for gauge fields in $dS_4/\bbZ_2$, with arbitrary parity-conserving interactions. In section \ref{sec:propagators}, we return to free fields, and present the boundary-to-bulk propagators for gauge fields in $dS_4/\bbZ_2$. In the scalar case, we find both the Neumann and Dirichlet propagators; for the $s>0$ gauge fields, we find the magnetic propagators, both as gauge potentials and as field strengths. 

In section \ref{sec:vasiliev}, we turn to Vasiliev gravity. We focus on the type-A and type-B versions of the theory, since the parity-violating versions cannot be defined on $dS_4/\bbZ_2$. We find the propagators for the zero-form master field in $dS_4/\bbZ_2$ with boundary conditions that preserve the higher-spin symmetry. We then plug these propagators into the $n$-point function calculations of \cite{Giombi:2010vg,Didenko:2012tv,Didenko:2013bj}, and show that the $n$-point functions all vanish. In section \ref{sec:discuss}, we conclude and discuss open questions.

\section{Preliminaries - geometry and gauge fields in de Sitter space} \label{sec:prelim_bulk}

\subsection{De Sitter space within $\bbR^{4,1}$ and the antipodal map} \label{sec:prelim_bulk:geometry}

We define de Sitter space $dS_4$ as the hyperboloid $x_\mu x^\mu = 1$ in the 4+1d Minkowski space $\bbR^{4,1}$. In this ``ambient formalism'', the de Sitter isometry group $SO(4,1)$ is identified with the rotation group in $\bbR^{4,1}$. We denote tensors in $\bbR^{4,1}$ with indices $(\mu,\nu,\dots)$, which are raised and lowered by the flat metric $\eta_{\mu\nu}$ with mostly-plus signature. The 3+1d tangent space to $dS_4$ at a point $x^\mu$ is picked out from the 4+1d vector space by the projector $P_\mu^\nu(x) = \delta_\mu^\nu - x_\mu x^\nu$. We use the same indices for tensors in $\bbR^{4,1}$ and $dS_4$, with the understanding that the latter are restricted to the span of $P_\mu^\nu(x)$. In this language, the intrinsic metric of $dS_4$ is $g_{\mu\nu}(x) = P_{\mu\nu}(x)$. Covariant derivatives in $dS_4$ are defined in terms of flat derivatives in $\bbR^{4,1}$ as:
\begin{align}
 \nabla_\mu v_\nu = P_\mu^\rho(x) P_\nu^\sigma(x) \del_\rho v_\sigma \ . \label{eq:tensor_covariant}
\end{align}
The d'Alembertian is defined as $\Box = \nabla_\mu\nabla^\mu$. The commutator of covariant derivatives takes the form:
\begin{align}
 [\nabla_\mu,\nabla_\nu]v^\rho = 2\delta_{[\mu}^\rho v_{\nu]} \ . \label{eq:commutator}
\end{align}

Every point $x^\mu \in dS_4$ has an antipodal point $-x^\mu$. The tangential projector $P_\mu^\nu(-x)$ is the same as $P_\mu^\nu(x)$, so that tensors at the two points can be directly compared. We say that a field $w_{\mu_1\dots\mu_k}(x)$ on $dS_4$ is antipodally even/odd when it goes into $+/-$ itself under the diffeomorphism $x\rightarrow -x$. In our tensor notation, this implies $w_{\mu_1\dots\mu_k}(-x) = \pm(-1)^k w_{\mu_1\dots\mu_k}(x)$ for antipodally even/odd fields. With this definition, the $dS_4$ metric and covariant derivative are antipodally even. 

The Levi-Civita tensor in $dS_4$ is obtained from the one in $\bbR^{4,1}$ through $\epsilon^{\mu\nu\rho\sigma} = \epsilon^{\mu\nu\rho\sigma\lambda}x_\lambda$. Under the antipodal map, $\epsilon^{\mu\nu\rho\sigma}$ flips sign. It follows that the antipodal map sends self-dual fields into anti-self-dual ones, and vice versa.

In \cite{Parikh:2002py}, it was argued that the antipodal map should involve a complex conjugation of dynamical fields, because a symmetry of the form $w_{\mu_1\dots\mu_k}(-x) = \pm w^*_{\mu_1\dots\mu_k}(x)$ ensures that antipodal points carry opposite charges. As noted in \cite{Neiman:2013hca}, this is incorrect: it is the symmetry \emph{without} complex conjugation that leads to opposite charges. Furthermore, the relation $w_{\mu_1\dots\mu_k}(-x) = \pm w^{\mu_1\dots\mu_k}(x)$ is invariant under internal symmetries of the form $w_{\mu_1\dots\mu_k}(x)\rightarrow e^{i\alpha}w_{\mu_1\dots\mu_k}(x)$, while the relation $w_{\mu_1\dots\mu_k}(-x) = \pm w^*_{\mu_1\dots\mu_k}(x)$ is not.

We conclude that in the standard C,P,T classification of discrete symmetries, the antipodal map in $dS_4$ is of the CT type. Indeed, the map interchanges past and future (hence the T), does not involve complex conjugation of fields (hence the C to revert the conjugation due to the T), and flips the spacetime orientation as captured by $\epsilon^{\mu\nu\rho\sigma}$ (hence no P that would revert the orientation flip due to the T). In fact, the antipodal map is CT in de Sitter space of any even spacetime dimension. In odd dimensions, the map is CPT, since the Levi-Civita tensor in that case is antipodally even. This distinction is contrary to the claim in \cite{Parikh:2002py} that the map is always CPT.

In addition to the real spacetime $dS_4$, we will make use of its complexification $dS_{4,\bbC}$. This is defined as the submanifold $x_\mu x^\mu = 1$ in the \emph{complex} space $\bbC^5$. Two slices of interest in $dS_{4,\bbC}$ are the imaginary past and future spaces:
\begin{align}
 \begin{split}
   H^- &= \left\{ x\in dS_{4,\bbC} \,| \Re x^\mu = 0\,,\ \Im x^0 < 0 \right\} \ ; \\ 
   H^+ &= \left\{ x\in dS_{4,\bbC} \,| \Re x^\mu = 0\,,\ \Im x^0 > 0 \right\} \ .
 \end{split} \label{eq:H}
\end{align}
The $H^\pm$ are 4d Euclidean anti-de Sitter (i.e. hyperbolic) spaces.

\subsection{Free gauge fields and field equations - tensor form} \label{sec:prelim_bulk:fields}

A spin-$s$ gauge field strength is a rank-$2s$ tensor $\varphi_{\mu_1\nu_1\mu_2\nu_2\dots\mu_s\nu_s}(x)$ that is antisymmetric in each pair of indices $\mu_k\nu_k$ and symmetric under the interchange of any two such pairs. In addition, all traces vanish, as does the antisymmetrization over any three indices. The cases $s=1,2$ correspond respectively to a Maxwell field strength $F_{\mu\nu}$ and a (linearized) Weyl tensor $C_{\mu_1\nu_1\mu_2\nu_2}$. As discussed in the Introduction, we consider an $m^2=2$ scalar as an ``honorary'' gauge field with $s=0$. For $s>0$, the field strength $\varphi_{\mu_1\nu_1\dots\mu_s\nu_s}$ decomposes into two pieces: one that is left-handed (anti-self-dual) in every $\mu_k\nu_k$ pair, and one that is right-handed (self-dual).

For $s=0$, the field $\varphi(x)$ satisfies the Klein-Gordon equation:
\begin{align}
 \Box\varphi - 2\varphi = 0 \ . \label{eq:scalar_diff}
\end{align}
This is the field equation $(\Box - R/6)\varphi = 0$ for a massless conformally coupled scalar (in our $dS_4$ space with unit radius, the Ricci scalar is $R = 12$).

For $s = 1$, we have the two free Maxwell equations:
\begin{align}
 \nabla^\mu\varphi_{\mu\nu} = 0 \quad ; \quad \nabla_{[\rho}\,\varphi_{\mu\nu]} = 0 \ . \label{eq:Maxwell_diff}
\end{align}

For $s\geq 2$, the analog of the first equation in \eqref{eq:Maxwell_diff} is sufficient:
\begin{align}
 \nabla^{\mu_1}\varphi_{\mu_1\nu_1\mu_2\nu_2\dots\mu_s\nu_s} = 0 \ . \label{eq:Weyl_diff}
\end{align}

To describe an interacting theory, the field strengths $\varphi_{\mu_1\nu_1\mu_2\nu_2\dots\mu_s\nu_s}(x)$ are not enough. Instead, one needs to work with gauge potentials $h_{\mu_1\mu_2\dots\mu_s}(x)$ \cite{Fronsdal:1978rb}. These are totally symmetric rank-$s$ tensors, which for $s\geq 4$ have a vanishing double trace: $h^{\nu\rho}_{\nu\rho\mu_5\dots\mu_s} = 0$. For $s=0$, we can define the ``potential'' $h(x)$ to coincide with the ``field strength'' $\varphi(x)$. For $s=1,2$, the potentials correspond respectively to a Maxwell potential $A_\mu$ and a metric perturbation $h_{\mu_1\mu_2}$. The free field equations for $h_{\mu_1\dots\mu_s}(x)$ in $dS_4$ take the form \cite{Fronsdal:1978vb}:
\begin{align}
 \begin{split}
   \Box h_{\mu_1\mu_2\dots\mu_s} &- s\nabla_{(\mu_1}\nabla^\nu h_{|\nu|\mu_2\dots\mu_s)} + \frac{s(s-1)}{2}\nabla_{(\mu_1}\nabla_{\mu_2}h^\nu_{|\nu|\mu_3\dots\mu_s)} \\
     &+ (s^2 - 2s - 2)h_{\mu_1\mu_2\dots\mu_s} + s(s-1)g_{(\mu_1\mu_2}h^\nu_{|\nu|\mu_3\dots\mu_s)} = 0 \ . \label{eq:h_diff}
 \end{split}
\end{align}
For $s=0$, this reduces to eq. \eqref{eq:scalar_diff}. For $s>0$, the field equations respect a gauge symmetry:
\begin{align}
 h_{\mu_1\mu_2\dots\mu_s} \rightarrow h_{\mu_1\mu_2\dots\mu_s} + \nabla_{(\mu_1}\theta_{\mu_2\dots\mu_s)} \ , \label{eq:gauge}
\end{align}
where the gauge parameter $\theta_{\mu_1\dots\mu_{s-1}}$ is a symmetric traceless tensor. There are enough degrees of freedom in $\theta_{\mu_1\dots\mu_{s-1}}$ to enforce transverse gauge $\nabla^{\mu_1}h_{\mu_1\mu_2\dots\mu_s} = 0$. For $s\geq 2$, one can use the remaining free initial data in $\theta_{\mu_1\dots\mu_{s-1}}$ to enforce traceless gauge $h^\nu_{\nu\mu_3\dots\mu_s} = 0$, which is then consistently evolved by the field equation \eqref{eq:h_diff} \cite{Mikhailov:2002bp}. In transverse traceless gauge, the field equation simplifies to:
\begin{align}
 \Box h_{\mu_1\dots\mu_s} + (s^2 - 2s - 2)h_{\mu_1\dots\mu_s} = 0 \ ; \quad \nabla^{\mu_1}h_{\mu_1\mu_2\dots\mu_s} = 0 \ ; \quad h^\nu_{\nu\mu_3\dots\mu_s} = 0 \ . \label{eq:h_diff_simple}
\end{align}
Some residual freedom in the gauge parameter $\theta_{\mu_1\dots\mu_{s-1}}$ still remains. In particular, at a point, any functional of $\theta_{\mu_1\dots\mu_{s-1}}$ with no contracted indices remains arbitrary.

With eqs. \eqref{eq:h_diff_simple} and the derivative commutator \eqref{eq:commutator}, one can reduce any expression in $h_{\mu_1\dots\mu_s}$ involving index contractions to an expression with 2 fewer derivatives. On the other hand, it's easy to see that with \emph{no} index contractions and fewer than $s$ derivatives, one cannot construct an expression that would be invariant under the residual gauge transformations. This brings us to the definition of the field strength $\varphi_{\mu_1\nu_1\dots\mu_s\nu_s}$ as the $s$-derivative gauge invariant:
\begin{align}
 \begin{split}
   \varphi_{\mu_1\nu_1\dots\mu_s\nu_s} = \nabla_{\mu_1}\dots\nabla_{\mu_s} h_{\nu_1\dots\nu_s} \quad \text{(}&\text{antisymmetrized over every }\mu_k\nu_k\text{ pair,} \\
   &\text{with all traces subtracted)} \ . \label{eq:phi_h}
 \end{split}
\end{align}
This definition coincides with the standard terminology for $s=1,2$, up to normalizations. In particular, if we take $A_\mu\equiv h_\mu$ and $h_{\mu\nu}$ to be the Maxwell potential and metric perturbation, then the Maxwell field strength and Weyl tensor read:
\begin{align}
 F_{\mu\nu} = 2\varphi_{\mu\nu} \quad ; \quad C_{\mu_1\nu_1\mu_2\nu_2} = -2\varphi_{\mu_1\nu_1\mu_2\nu_2} \ . \label{eq:mismatch}
\end{align}
In the definition \eqref{eq:phi_h}, the $s$ covariant derivatives are effectively symmetrized, since any derivative commutators yield trace pieces via \eqref{eq:commutator}. The correct index symmetries for a field strength directly follow. $\varphi_{\mu_1\nu_1\dots\mu_s\nu_s}$ is gauge-invariant, as can be seen by plugging the gauge variation \eqref{eq:gauge} into its definition. Indeed, the $\mu_k\nu_k$ antisymmetrizations and the derivative in \eqref{eq:gauge} reduce the gauge variation to derivative commutators, which again become trace pieces due to \eqref{eq:commutator}. 

Finally, when $h_{\mu_1\dots\mu_s}$ satisfies the field equation \eqref{eq:h_diff}, we get that $\varphi_{\mu_1\nu_1\dots\mu_s\nu_s}$ satisfies the field equation \eqref{eq:Weyl_diff}. This is easiest to see in transverse traceless gauge. Recall that under eqs. \eqref{eq:h_diff_simple}, any expression in $h_{\mu_1\dots\mu_s}$ with index contractions can be reduced to an expression with fewer derivatives. In particular, $\nabla^{\mu_1}\varphi_{\mu_1\nu_1\dots\mu_s\nu_s}$ reduces to an expression with fewer than $s$ derivatives. But we've seen that under eqs. \eqref{eq:h_diff_simple}, any gauge invariant with fewer than $s$ derivatives must vanish. The field equation \eqref{eq:Weyl_diff} is thus established.

\subsection{Spinors and twistors in $dS_4$} \label{sec:prelim_bulk:spinors}

The de Sitter group $SO(4,1)$ has a unique spin-1/2 representation, with 4-component Dirac spinors. This is the \emph{twistor space} \cite{Penrose:1986ca,Ward:1990vs} of $dS_4$, though we will mostly use the word ``spinor'', in keeping with the $\bbR^{4,1}$ perspective. We use indices $(a,b,\dots)$ for the $SO(4,1)$ spinors. The spinor space has a symplectic metric $I_{ab}$, which is used to raise and lower indices via $\psi_a = I_{ab}\psi^b$ and $\psi^a = \psi_b I^{ba}$, where $I^{ac}I_{bc} = \delta^a_b$. Tensor and spinor indices are related through the gamma matrices $(\gamma_\mu)^a{}_b$, which satisfy the Clifford algebra $\{\gamma_\mu,\gamma_\nu\} = -2\eta_{\mu\nu}$. These 4+1d gamma matrices can be realized as the usual 3+1d ones, with the addition of $\gamma_5$ (in our notation, $\gamma_4$) for the fifth direction in $\bbR^{4,1}$. These matrices can be represented in $2\times 2$ block notation as:
\begin{align}
 \begin{split}
   I_{ab} &= -i\begin{pmatrix} 0 & \sigma_2 \\ \sigma_2 & 0 \end{pmatrix} \ ; \\
   (\gamma^0)^a{}_b &= \begin{pmatrix} 0 & 1 \\ 1 & 0 \end{pmatrix} \ ; \quad 
   (\gamma^k)^a{}_b = -i\begin{pmatrix} \sigma^k & 0 \\ 0 & -\sigma^k \end{pmatrix} \ ; \quad 
   (\gamma^4)^a{}_b = \begin{pmatrix} 0 & -1 \\ 1 & 0 \end{pmatrix} \ , \label{eq:gamma}
 \end{split}
\end{align}
where $\sigma^k$ with $k=1,2,3$ are the Pauli matrices. The $\gamma^\mu_{ab}$ are antisymmetric and traceless in their spinor indices. We define the antisymmetric product of gamma matrices as:
\begin{align}
 \gamma^{\mu\nu}_{ab} \equiv \gamma^{[\mu}_{ac}\gamma^{\nu]c}{}_b \ .
\end{align}
The $\gamma^{\mu\nu}_{ab}$ are symmetric in their spinor indices. Useful identities include: 
\begin{align}
 \begin{split}
   &\gamma^\mu_{ab}\gamma_\nu^{ab} = -4\delta^\mu_\nu \ ; \quad \gamma^{\mu\nu}_{ab}\gamma_{\rho\sigma}^{ab} = 8\delta^{[\mu}_{[\rho}\delta^{\nu]}_{\sigma]} \ ; \quad
   \gamma_\mu^{ab}\gamma^\mu_{cd} = I^{ab}I_{cd} - 4\delta^{[a}_{[c} \delta^{b]}_{d]} \ ; \\
   &\epsilon^{abcd} = -3I^{[ab}I^{cd]} \ ; \quad \epsilon^{abcd}I_{cd} = -2I^{ab} \ ; \quad \epsilon^{abcd}\gamma^\mu_{cd} = 2\gamma^{\mu ab} \ .
 \end{split}
\end{align}
We can use $\gamma_\mu^{ab}$ to convert between 4+1d vectors and traceless bispinors as:
\begin{align}
 v^{ab} = \gamma_\mu^{ab}v^\mu \ ; \quad v^\mu = -\frac{1}{4}\gamma^\mu_{ab}v^{ab} \ ; \quad u\cdot v \equiv u_\mu v^\mu = -\frac{1}{4}u_{ab} v^{ab} \ , \label{eq:conversion_5d}
\end{align}
Similarly, we can use $\gamma_{\mu\nu}^{ab}$ to convert between bivectors and symmetric spinor matrices:
\begin{align}
 f^{ab} = \frac{1}{2}\gamma_{\mu\nu}^{ab}f^{\mu\nu} \ ; \quad f^{\mu\nu} = \frac{1}{4}\gamma^{\mu\nu}_{ab} f^{ab} \ .
\end{align}
Further details may be found in \cite{Neiman:2013hca}. 

When we choose a point $x\in dS_4$, the Dirac representation of $SO(4,1)$ becomes identified with the Dirac representation of the Lorentz group $SO(3,1)$ at $x$. It then decomposes into left-handed and right-handed Weyl representations. The decomposition is accomplished by the pair of projectors:
\begin{align}
 \begin{split}
   P_L{}^a{}_b(x) &= \frac{1}{2}\left(\delta^a_b - ix^\mu\gamma_\mu{}^a{}_b \right) = \frac{1}{2}\left(\delta^a_b - ix^a{}_b \right) \ ; \\
   P_R{}^a{}_b(x) &= \frac{1}{2}\left(\delta^a_b + ix^\mu\gamma_\mu{}^a{}_b \right) = \frac{1}{2}\left(\delta^a_b + ix^a{}_b \right) \ . \label{eq:projectors}
 \end{split}
\end{align}
These serve as an $x$-dependent version of the familiar chiral projectors in $\bbR^{3,1}$. Given an $SO(4,1)$ spinor $\psi^a$, we denote its left-handed and right-handed components at $x$ as $\psi_{L/R}^a(x) = (P_{L/R}){}^a{}_b(x)\psi^b$. As in our treatment of tensors, it is possible to use the $(a,b,\dots)$ indices for both $SO(4,1)$ and $SO(3,1)$ Dirac spinors. In addition, at a point $x\in dS_4$, it will be convenient to use left-handed $(\alpha,\beta,\dots)$ and right-handed $(\dot\alpha,\dot\beta,\dots)$ Weyl indices, which are taken to imply $P_L(x)$ and $P_R(x)$ projections, respectively. Thus, for a Dirac spinor $\psi^a$, we have the projections $\psi_L^\alpha(x)$ and $\psi_R^{\dot\alpha}(x)$. In this scheme, the matrices $P^L_{ab}(x)$ and $P^R_{ab}(x)$ serve as the spinor metrics $\epsilon_{\alpha\beta}$ and $\epsilon_{\dot\alpha\dot\beta}$ for the two Weyl spinor spaces.

For a vector $v^\mu$ in the 3+1d tangent space at a de Sitter point $x$, the bispinor $v^{ab}$ can be decomposed into Weyl components $v^{\alpha\dot\beta} = -v^{\dot\beta\alpha}$. For such vectors, we therefore have:
\begin{align}
 v^{\alpha\dot\alpha} = \gamma_\mu^{\alpha\dot\alpha}v^\mu \ ; \quad v^\mu = -\frac{1}{2}\gamma^\mu_{\alpha\dot\alpha}v^{\alpha\dot\alpha} \ ;
 \quad u_\mu v^\mu = -\frac{1}{2}u_{\alpha\dot\alpha}v^{\alpha\dot\alpha} \ . \label{eq:conversion_4d}
\end{align}

The power of this formalism is that the $SO(4,1)$ spinors are flat, just like the $SO(4,1)$ vectors. We can therefore transport them freely from one de Sitter point to another. What changes from point to point is the spinor's decomposition into left-handed and right-handed parts. As a special case, the identity $P_R^{ab}(-x) = P_L^{ab}(x)$ defines an isomorphism between left-handed spinors at $x$ and right-handed spinors at $-x$. This is consistent with the fact that self-duality signs get flipped by the antipodal map.

Covariant derivatives for Weyl spinors in $dS_4$ can be constructed from the 4+1d flat derivative, in analogy with the tensor formula \eqref{eq:tensor_covariant}:
\begin{align}
 \nabla_{\alpha\dot\alpha}\psi^b_{L/R}(x) = (P_{L/R})^b{}_c(x)\,\del_{\alpha\dot\alpha}\psi_{L/R}^c(x) \ . \label{eq:spinor_covariant}
\end{align}

\subsection{Free gauge fields and field equations - spinor form} \label{sec:prelim_bulk:spinor_fields}

A field strength tensor with the appropriate index symmetries can be translated into a totally symmetric rank-$2s$ spinor as:
\begin{align}
 \varphi_{\mu_1\nu_1\dots\mu_s\nu_s}(x) ={}& \frac{1}{4^s}\,\gamma_{\mu_1\nu_1}^{a_1 b_1}\dots \gamma_{\mu_s\nu_s}^{a_s b_s}\, \varphi_{a_1 b_1\dots a_s b_s}(x) \ ,
\end{align}
where the only non-vanishing components of $\varphi_{a_1\dots a_{2s}}$ are the totally left-handed $\varphi_{\alpha_1\dots\alpha_{2s}}$ and the totally right-handed $\varphi_{\dot\alpha_1\dots\dot\alpha_{2s}}$. For $s>0$, the field-strength spinors satisfy the field equations:
\begin{align}
 \nabla^{\alpha_1\dot\beta}\varphi_{\alpha_1\alpha_2\dots\alpha_{2s}}(x) = 0 \quad ; \quad 
 \nabla^{\beta\dot\alpha_1}\varphi_{\dot\alpha_1\dot\alpha_2\dots\dot\alpha_{2s}}(x) = 0 \ . \label{eq:spinor_diff}
\end{align}

A gauge potential $h_{\mu_1\dots\mu_s}$ in traceless gauge can be translated into spinor form as:
\begin{align}
 h_{\mu_1\dots\mu_s}(x) = \frac{(-1)^s}{2^s}\, \gamma_{\mu_1}^{\alpha_1\dot\alpha_1}\dots\gamma_{\mu_s}^{\alpha_s\dot\alpha_s}
  h_{\alpha_1\dots\alpha_s\dot\alpha_1\dots\dot\alpha_s}(x) \ , 
\end{align}
where $h_{\alpha_1\dots\alpha_s\dot\alpha_1\dots\dot\alpha_s}$ is symmetric in both its dotted and undotted indices. We will not consider here the extension to half-integer spins. The field equations and gauge conditions \eqref{eq:h_diff_simple} translate directly into spinor language. On the other hand, the relation  \eqref{eq:phi_h} between potentials and field strengths simplifies considerably. It can be formulated succinctly in spinor language as:
\begin{align}
 \begin{split}
   \varphi_{\alpha_1\beta_1\dots\alpha_s\beta_s} &= \frac{1}{2^s}\nabla_{(\alpha_1}{}^{\dot\alpha_1}\dots\nabla_{\alpha_s}{}^{\dot\alpha_s}
     h_{\beta_1\dots\beta_s)\dot\alpha_1\dots\dot\alpha_s} \ ; \\
   \varphi_{\dot\alpha_1\dot\beta_1\dots\dot\alpha_s\dot\beta_s} &= \frac{1}{2^s}\nabla^{\alpha_1}{}_{(\dot\alpha_1}\dots\nabla^{\alpha_s}{}_{\dot\alpha_s}
     h_{|\alpha_1\dots\alpha_s|\dot\beta_1\dots\beta_s)} \ .
 \end{split} \label{eq:phi_h_spinor}
\end{align} 

\section{Preliminaries: geometry and gauge fields at $\scri^\pm$} \label{sec:prelim_scri}

In this section, we outline the asymptotic geometry of $dS_4$, along with the appropriate boundary data for the gauge fields. For the latter, we will use the conformal properties of the field equations \eqref{eq:scalar_diff}-\eqref{eq:Weyl_diff}.

\subsection{Asymptotic geometry} \label{sec:prelim_scri:geometry}

The asymptotic boundary of $dS_4$ is a pair of spacelike conformal 3-spheres - one in the infinite past ($\scri^-$), and the other in the infinite future ($\scri^+$).
In the flat 4+1d picture, these can be viewed as the 3-spheres of past-pointing and future-pointing null directions in $\bbR^{4,1}$. The antipodal map interchanges $\scri^-$ and $\scri^+$, in such a way that the lightcone of a point on $\scri^-$ refocuses at the antipodal point on $\scri^+$. A bulk point $x$ is said to ``approach infinity'' when the unit vector $x^\mu$ is highly boosted, i.e. when its components become very large. This condition is not invariant under the de Sitter group $SO(4,1)$, but that is to be expected: the statement that a point is ``very far away'' cannot be invariant under large translations.

$\scri^-$ and $\scri^+$ can be assigned an orientation by contracting the bulk Levi-Civita tensor $\epsilon^{\mu\nu\rho\sigma}$ with the future-pointing or past-pointing timelike normal $\pm n^\mu$ (where we take $n^\mu$ to be the future-pointing choice). To avoid a preferred global time direction, we must use $-n^\mu$ at $\scri^-$ and $+n^\mu$ at $\scri^+$, or vice versa. This choice of normals is antipodally even, while $\epsilon^{\mu\nu\rho\sigma}$ is antipodally odd. Therefore, in this scheme, the antipodal map reverses the orientation of $\scri^\pm$. In this sense, $\scri^-$ and $\scri^+$ have opposite orientations.

To include $\scri^\pm$ in the spacetime manifold, we perform a conformal completion: we choose a time coordinate $z$ that vanishes on $\scri^\pm$, such that the conformally rescaled metric $z^2 g_{\mu\nu}$ is regular at $z=0$. We can then define a metric on $\scri^\pm$ as:
\begin{align}
 q_{\mu\nu}(x) = \lim_{z\rightarrow 0} z^2 g_{\mu\nu}(x) \quad \text{ (pulled back to }z=0\text{)} \ . \label{eq:q}
\end{align}
Since we are free to multiply $z$ by any function of the spatial coordinates, the metric \eqref{eq:q} is only defined conformally. To remove any ambiguity between $\scri^-$ and $\scri^+$, we choose the $z$ coordinate to be antipodally odd, such that $\scri^-$ and $\scri^+$ correspond to $z\rightarrow 0^-$ and $z\rightarrow 0^+$, respectively. 

\subsection{Boundary data for the scalar field ($s=0$)}

We now turn to the issue of appropriate boundary data for our fields on $\scri^\pm$. We begin with the $m^2=2$ scalar $\varphi(x)$, satisfying the field equation \eqref{eq:scalar_diff}. As already mentioned, eq. \eqref{eq:scalar_diff} can be written as:
\begin{align}
 \Box\varphi - \frac{1}{6}R\varphi = 0 \ , \label{eq:conformal_scalar}
\end{align}
where $R$ is the Ricci scalar. Eq. \eqref{eq:conformal_scalar} is invariant under the conformal rescaling $g_{\mu\nu}\rightarrow z^2 g_{\mu\nu}$, where $\varphi$ has conformal weight $1$ (we say that a quantity has conformal weight $\Delta$ if it scales as $z^{-\Delta}$ under $g_{\mu\nu}\rightarrow z^2 g_{\mu\nu}$). Since the metric $z^2 g_{\mu\nu}$ is regular at $\scri^\pm$, we conclude that the rescaled field $z^{-1}\varphi$ can be Cauchy-evolved from $\scri^\pm$. We can therefore define configuration and momentum fields on $\scri^\pm$ as:
\begin{align}
 \left.\phi(x)\right|_{\scri^\pm} \equiv \pm\lim_{z\rightarrow 0^\pm} \frac{\varphi(x)}{z} \quad ; \quad 
 \left.\pi(x)\right|_{\scri^\pm} \equiv \lim_{z\rightarrow 0^\pm} \ddsimple{}{z}\left(\frac{\varphi(x)}{z}\right) \ . \label{eq:phi_pi}
\end{align}
See \cite{Anninos:2012ft} for the dS/CFT perspective on these definitions. The chosen sign factors ensure that an antipodally even/odd $\varphi(x)$ induces the same symmetry on $\phi(x)$ and $\pi(x)$. As fields on $\scri^\pm$, $\phi(x)$ and $\pi(x)$ have respective conformal weights $1$ and $2$ under rescalings of the metric \eqref{eq:q}. The weights add up to $3$, as appropriate for canonical conjugates in a 3-dimensional CFT. A solution to the field equation \eqref{eq:scalar_diff} is uniquely determined by the boundary data $\{\phi(x),\pi(x)\}$ on e.g. $\scri^-$.

\subsection{Boundary data for gauge fields ($s\geq 1$)}

We now turn to the gauge field strengths $\varphi_{\mu_1\nu_1\dots\mu_s\nu_s}(x)$ with field equations \eqref{eq:Maxwell_diff}-\eqref{eq:Weyl_diff}. On spatial slices of $dS_4$, for which $\scri^\pm$ are limiting cases, we can decompose $\varphi_{\mu_1\nu_1\dots\mu_s\nu_s}$ into electric and magnetic components with respect to the future-pointing unit normal $n^\mu$. A-priori, every $\mu_k\nu_k$ index pair can be decomposed separately. However, due to the index symmetries of $\varphi_{\mu_1\nu_1\dots\mu_s\nu_s}$, a simultaneous Hodge dual on any two pairs yields the original field with a minus sign:
\begin{align}
 \frac{1}{4}\epsilon_{\mu_1\nu_1}{}^{\rho_1\sigma_1}\epsilon_{\mu_2\nu_2}{}^{\rho_2\sigma_2}\varphi_{\rho_1\sigma_1\rho_2\sigma_2\mu_3\nu_3\dots\mu_s\nu_s}
  = -\varphi_{\mu_1\nu_1\mu_2\nu_2\mu_3\nu_3\dots\mu_s\nu_s} \ .
\end{align}
Thus, pieces of $\varphi_{\mu_1\nu_1\dots\mu_s\nu_s}$ with an even (odd) number of magnetic $\mu_k\nu_k$ pairs are all equivalent to the piece with zero (one) such pairs. We can therefore decompose $\varphi_{\mu_1\nu_1\dots\mu_s\nu_s}$ into electric and magnetic parts as follows:
\begin{align}
 \begin{split}
   E_{\mu_1\mu_2\dots\mu_s} &\equiv n^{\nu_1}n^{\nu_2}\dots n^{\nu_s} \varphi_{\mu_1\nu_1\mu_2\nu_2\dots\mu_s\nu_s} \ ; \\
   B_{\mu_1\mu_2\dots\mu_s} &\equiv \frac{1}{2}\epsilon_{\mu_1\nu_1}{}^{\rho\sigma}n^{\nu_1}n^{\nu_2}\dots n^{\nu_s} \varphi_{\rho\sigma\mu_2\nu_2\dots\mu_s\nu_s} \ .
 \end{split} \label{eq:E_B}
\end{align}
The tensors \eqref{eq:E_B} are purely spatial, totally symmetric and traceless.

On a spatial slice, the field equations \eqref{eq:Maxwell_diff}-\eqref{eq:Weyl_diff} decompose into constraints and dynamical equations. The constraint equations read:
\begin{align}
 D_{\mu_1}E^{\mu_1\mu_2\dots\mu_s} = D_{\mu_1}B^{\mu_1\mu_2\dots\mu_s} = 0 \ , \label{eq:div_E_B}
\end{align}
where $D_\mu$ is the spatial covariant derivative. The dynamical equations evolve the $\{B_{\mu_1\dots\mu_s}(x),E_{\mu_1\dots\mu_s}(x)\}$ values on a spatial slice into a spacetime solution.

To find the appropriate boundary data on $\scri^\pm$, we note that the field equations \eqref{eq:Maxwell_diff}-\eqref{eq:Weyl_diff} are conformally invariant, with conformal weight $1-s$ for $\varphi_{\mu_1\nu_1\dots\mu_s\nu_s}$. We then read off from \eqref{eq:E_B} that both $E_{\mu_1\mu_2\dots\mu_s}$ and $B_{\mu_1\mu_2\dots\mu_s}$ have conformal weight $1$ (note that $n^\mu$ and $\epsilon^{\mu\nu\rho\sigma}$ have weights $1$ and $4$, respectively). Since the rescaled metric $z^2 g_{\mu\nu}$ is regular at $\scri^\pm$, we conclude that the proper boundary data is given by:
\begin{align}
 \begin{split}
   \left.\calE_{\mu_1\dots\mu_s}(x)\right|_{\scri^\pm} &\equiv (\pm 1)^{s+1} \lim_{z\rightarrow 0^\pm}\frac{E_{\mu_1\dots\mu_s}(x)}{z} \ ; \\
   \left.\calB_{\mu_1\dots\mu_s}(x)\right|_{\scri^\pm} &\equiv (\pm 1)^s \lim_{z\rightarrow 0^\pm}\frac{B_{\mu_1\dots\mu_s}(x)}{z} \ .
 \end{split} \label{eq:cal_E_B}
\end{align}
The extra sign factors compensate for the fact that $n^\mu$ and $\epsilon^{\mu\nu\rho\sigma}$ are antipodally odd. With the definition \eqref{eq:cal_E_B}, an antipodally even/odd $\varphi_{\mu_1\nu_1\dots\mu_s\nu_s}$ induces the same symmetry on $\calE_{\mu_1\dots\mu_s}$ and $\calB_{\mu_1\dots\mu_s}$.

Under the conformal rescaling $g_{\mu\nu}\rightarrow z^2 g_{\mu\nu}$, the constraints \eqref{eq:div_E_B} become:
\begin{align}
 \calD_{\mu_1}\calE^{\mu_1\mu_2\dots\mu_s} = \calD_{\mu_1}\calB^{\mu_1\mu_2\dots\mu_s} = 0 \ , \label{eq:div_E_B_scri}
\end{align}
where $\calD_\mu$ is the covariant derivative for the metric \eqref{eq:q} at $\scri^\pm$. A solution to the field equations \eqref{eq:Maxwell_diff}-\eqref{eq:Weyl_diff} is uniquely determined by the boundary data $\{\calB_{\mu_1\dots\mu_s}(x),\calE_{\mu_1\dots\mu_s}(x)\}$ on e.g. $\scri^-$, subject to the constraints \eqref{eq:div_E_B_scri}.

\section{Free fields: antipodal symmetry and asymptotics} \label{sec:relations}

\subsection{Results}

In this section, we prove the following results, which relate the antipodal symmetry of free gauge fields with their asymptotic behavior:
\begin{theorem}[Free scalar fields] \label{thm:free_scalar}
 Consider a free scalar field in $dS_4$, satisfying the field equation $(\Box - 2)\varphi = 0$. Then the space of solutions is a direct sum of two subspaces:
 \begin{enumerate}[leftmargin=*]
  \item Antipodally even solutions, which satisfy vanishing Dirichlet conditions $\phi(x) = 0$ on $\scri$;
  \item Antipodally odd solutions, which satisfy vanishing Neumann conditions $\pi(x) = 0$ on $\scri$.
 \end{enumerate}
\end{theorem}
\begin{theorem}[Free gauge fields] \label{thm:free_gauge}
 Consider a free spin-$s$ gauge field in $dS_4$, satisfying the field equations \eqref{eq:Maxwell_diff}-\eqref{eq:Weyl_diff}. Then the space of solutions is a direct sum of two subspaces:
 \begin{enumerate}[leftmargin=*]
  \item Antipodally even solutions, which are purely magnetic on $\scri$, i.e. satisfy $\calE_{\mu_1\dots\mu_s}(x) = 0$;
  \item Antipodally odd solutions, which are purely electric on $\scri$, i.e. satisfy $\calB_{\mu_1\dots\mu_s}(x) = 0$.
 \end{enumerate}
\end{theorem}
With regard to Theorem \ref{thm:free_gauge}, we note that an antipodally even/odd field strength can always be derived from a gauge potential with the same symmetry: start with any gauge potential for the given field strength, and take its antipodally even/odd piece; the remaining piece is necessarily pure gauge.

Theorems \ref{thm:free_scalar}-\ref{thm:free_gauge} are not surprising, once one realizes that the fields in question propagate along lightrays. Since the lightcone from a point on $\scri^-$ refocuses at its antipode on $\scri^+$, it is natural for the boundary data on $\scri^\pm$ to be antipodally symmetric. What remains to be established is the sign of this antipodal symmetry, which depends on the type of boundary data.

Once we know the antipodal symmetry of each type of boundary data, it is easy to demonstrate the \emph{bulk} antipodal symmetry of solutions with only this type of boundary data non-vanishing. Indeed, since the field equations are antipodally symmetric, the solution's antipodal image is also a solution. But this will have the same boundary data as the original solution on e.g. $\scri^-$, up to an overall sign. The uniqueness of Cauchy evolution then implies that the solution must coincide (up to sign) with its antipodal image.

It remains, then, to map each type of boundary data to its antipodal symmetry on $\scri^\pm$. In section \ref{sec:relations:proof_from_smoothness}, we accomplish this by assuming smoothness through $\scri^\pm$ on conformally compactified $dS_4$. We will justify this assumption through explicit solutions in sections \ref{sec:relations:scalar_solutions}-\ref{sec:relations:gauge_solutions}.

\subsection{Proof from smoothness on conformally compactified $dS_4$} \label{sec:relations:proof_from_smoothness}

We've seen that in order to prove theorems \ref{thm:free_scalar}-\ref{thm:free_gauge}, it is enough to map each type of boundary data to its antipodal symmetry on $\scri^\pm$. We will now do this, using the conformal symmetry of the field equations. The conformal group $SO(4,2)$ in 3+1d spacetime can be realized as the group of rotations in 4+2d flat space. In this realization, our $dS_4$ is a section of the lightcone in $\bbR^{4,2}$. Specifically, a point $x\in dS_4$, which in $\bbR^{4,1}$ is described by the unit spacelike vector $x^\mu$, is associated with the following null vector in $\bbR^{4,2}$:
\begin{align}
 dS_4: \quad X^A = (1,x^\mu) \ , \label{eq:X}
\end{align}
where the index $A$ takes the values $A=-1,0,1,2,3,4$, and the new $A=-1$ direction is timelike. Other sections of the lightcone describe conformally related 3+1d metrics. Forgetting the particular section \eqref{eq:X}, one can define conformally compactified $dS_4$ as the projective lightcone in $\bbR^{4,2}$, i.e. the space of nonzero null $X^A$ modulo rescalings $X^A\rightarrow zX^A$. 

Fields on $dS_4$ can be written as functions of the null vector $\ell^A$. Fields with conformal weight $\Delta$ will scale as $z^{-\Delta}$ under $X^A \rightarrow zX^A$.
The asymptotic 3-sphere $\scri^-$ gets mapped into an ordinary 3d surface on the projective lightcone in $\bbR^{4,2}$. Crucially, $\scri^+$ gets mapped into the \emph{same} surface, so that antipodal points on $\scri^\pm$ (but not elsewhere) become identified. The $\scri^\pm$ become singular and distinct from one another only in the particular section \eqref{eq:X}. After rescaling $X^A\rightarrow zX^A$, with $z$ the time coordinate from section \ref{sec:prelim_scri}, the points of $\scri^\pm$ take the form:
\begin{align}
 \scri^\pm: \quad L^A = (0^\pm, \ell^\mu) \ , \label{eq:L}
\end{align}
where the $\ell^\mu$ are future-pointing null vectors forming a section of the lightcone in $\bbR^{4,1}$.

Now, consider a solution $\varphi(x)$ to the free scalar equation \eqref{eq:scalar_diff} in $dS_4$. Assume that in conformally compactified $dS_4$, this solution is regular on the surface corresponding to $\scri^\pm$. Taking into account the rescaling between the conformal frames \eqref{eq:X}-\eqref{eq:L}, this means that $\varphi(x)/z$ is regular on the section \eqref{eq:L}. Now, as we've seen, the antipodal map between the $\scri^\pm$ is trivial on the section \eqref{eq:L}. This implies that $\varphi(x)/z$, as well as $\del_z(\varphi(x)/z)$, are antipodally even. The antipodal symmetries in Theorem \ref{thm:free_scalar} can now be read off from the definitions \eqref{eq:phi_pi}.

Similarly, consider a free field-strength solution $\varphi_{\mu_1\nu_1\dots\mu_s\nu_s}(x)$ on $dS_4$ with $s\geq 1$. Assume again that in conformally compactified $dS_4$, the solution is regular on the surface corresponding to $\scri^\pm$, i.e. on the section \eqref{eq:L}. To draw conclusions about our tensor field in $dS_4$, we must take its components with respect to directions that are smooth through \eqref{eq:L}. The directions tangential to $\scri^\pm$ with this property are antipodally even, but the normal direction (e.g. the future-pointing one) is antipodally odd. Taking into account the appropriate conformal weight, we conclude that components of  $z^{s-1}\varphi_{\mu_1\nu_1\dots\mu_s\nu_s}(x)$ with an even/odd number of normal indices are antipodally even/odd. The antipodal symmetries in Theorem \ref{thm:free_gauge} can now be read off from the definitions \eqref{eq:E_B},\eqref{eq:cal_E_B}.

It remains to justify the assumption of smoothness through $\scri^\pm$ on conformally compactified $dS_4$. In the next subsections, we show that this property is indeed satisfied by a spanning set of solutions to the free field equations. 

\subsection{Smooth solutions on conformally compactified $dS_4$: scalar field} \label{sec:relations:scalar_solutions}

We begin with the scalar case. Our solutions for $\varphi(x)$ will be parametrized by a point $\xi$ in the imaginary future slice $H^+$ or the imaginary past slice $H^-$ of complexified de Sitter space; see eq. \eqref{eq:H}. In the 4+1d language, $\xi$ is encoded by an imaginary timelike vector $\xi^\mu$ such that $\xi_\mu\xi^\mu = 1$ and $\Im\xi^0 \gtrless 0$, respectively. We then consider the solution:
\begin{align}
 \varphi(x) = \frac{1}{x\cdot\xi - 1} \ , \label{eq:scalar_solution}
\end{align}
where $x\cdot\xi \equiv x_\mu\xi^\mu$. This is just a bulk 2-point function between the points $x$ and $\xi$ \cite{Fronsdal:1974ew}. The denominator in \eqref{eq:scalar_solution} is proportional to the squared distance between $x^\mu$ and $\xi^\mu$ in the complexified $\bbR^{4,1}$:
\begin{align}
 x_\mu\xi^\mu - 1 = -\frac{1}{2}(x_\mu - \xi_\mu)(x^\mu - \xi^\mu) \ .
\end{align}
Since lightrays in $dS_4$ are also lightrays in $\bbR^{4,1}$, the solution \eqref{eq:scalar_solution} is singular along the lightcone of $\xi$. For $\xi\in H^\pm$, this lightcone intersects neither the real spacetime $dS_4$ nor the opposite imaginary slice $H^\mp$. The solution is therefore regular on $dS_4$ and $H^\mp$. The regularity on $H^\mp$ means that the solution has respectively positive/negative frequency in the Bunch-Davies sense. 

It is easy to show that the solutions \eqref{eq:scalar_solution} form a spanning set. First, note that the positive-frequency and negative-frequency solutions to the field equation \eqref{eq:scalar_diff} form irreducible representations of the de Sitter group $SO(4,1)$. We have seen that the solutions \eqref{eq:scalar_solution} with $\xi\in H^\pm$ belong to these representations. On the other hand, these solutions also \emph{span} representations of $SO(4,1)$: an $SO(4,1)$ rotation yields another solution of the form \eqref{eq:scalar_solution}, with a rotated value for $\xi^\mu$. These representations must then coincide with the positive/negative frequency representations, since the latter are irreducible. We conclude that the solutions \eqref{eq:scalar_solution} span the full solution space of the field equation \eqref{eq:scalar_diff}.

Finally, it's easy to see that the solutions \eqref{eq:scalar_solution} are regular on $\scri^\pm$ in conformally compactified $dS_4$. This follows directly from the fact that $\scri^\pm$ doesn't lie on the lightcone of the imaginary point $\xi$. Explicitly, the solutions can be written in the 4+2d language as:
\begin{align}
 \varphi(X) = \frac{1}{X\cdot\Xi} \ ,
\end{align}
where $X^A = (1,x^\mu)$ and $\Xi^A = (1,\xi^\mu)$. The solutions are now manifestly $SO(4,2)$-covariant with conformal weight $1$, and regular on the asymptotic section \eqref{eq:L}.

\subsection{Smooth solutions on conformally compactified $dS_4$: gauge fields} \label{sec:relations:gauge_solutions}

We now turn to spin-$s$ gauge field strengths with $s\geq 1$. We will again use 2-point functions between the measurement point $x\in dS_4$ and a point $\xi\in H^\pm$ in the imaginary future/past. To write the solutions compactly, we will use the spinor language of sections \ref{sec:prelim_bulk:spinors}-\ref{sec:prelim_bulk:spinor_fields}. Our 2-point functions will have opposite handedness at $x$ and $\xi$. In principle, they can be derived from the bulk-to-bulk gauge-potential propagators, given e.g. in \cite{Costa:2014kfa}. However, the end result is much simpler than the calculation, so we present the field-strength solutions directly.

To encode the polarization, we will use a Weyl spinor $M$ at the point $\xi$. For e.g. left-handed solutions at $x$, we will take $M$ to be right-handed at $\xi$. In the 4+1d language, $M$ is then encoded by a 4-component Dirac spinor $M^a$, subject to the condition $P_L{}^a{}_b(\xi)M^b = 0$. Our left-handed solutions then read:
\begin{align}
 \varphi_{\alpha_1\alpha_2\dots\alpha_{2s}}(x) = \frac{M^L_{\alpha_1}(x)M^L_{\alpha_2}(x)\dots M^L_{\alpha_{2s}}(x)}{(x\cdot\xi - 1)^{2s+1}} \ , \label{eq:spinor_solution}
\end{align}
where $M_L^\alpha(x)$ is the projection $P_L{}^\alpha{}_b(x) M^b$ of $M^a$ onto the left-handed spinor space at $x$. The scalar solution \eqref{eq:scalar_solution} is contained in \eqref{eq:spinor_solution} as the $s=0$ case. One can verify that \eqref{eq:spinor_solution} indeed solves the field equation \eqref{eq:spinor_diff}, using the relations:
\begin{align}
 \begin{split}
   M_L^\alpha(x)\nabla_{\alpha\dot\alpha}(x\cdot\xi) &= -i(x\cdot\xi - 1) M^R_{\dot\alpha}(x) \ ; \\ 
   \nabla_{\alpha\dot\alpha} M_L^\alpha(x) &= -2i M^R_{\dot\alpha}(x) \ ; \\
   M_L^\alpha(x)\nabla_{\alpha\dot\alpha} M_L^\beta(x) &= -iM^R_{\dot\alpha}(x) M_L^\beta(x) \ .
 \end{split} \label{eq:diff_M_L_spinor}
\end{align} 
Field strengths with the opposite handedness can be obtained by interchanging the $P_L$ and $P_R$ projectors, or, equivalently, by substituting $x\rightarrow -x$. As in the scalar case, the solutions \eqref{eq:spinor_solution} with $\xi\in H^\pm$ have positive/negative frequency in the Bunch-Davies sense. It then follows similarly that the solutions \eqref{eq:spinor_solution}, along with their right-handed counterparts, span the full solution space of the field equation \eqref{eq:spinor_diff}. 

Finally, the solutions \eqref{eq:spinor_solution} are again regular on $\scri^\pm$ in conformally compactified $dS_4$, because $\scri^\pm$ doesn't lie on the lightcone of $\xi$. To see this explicitly in 4+2d language, we must introduce the spinors of $\bbR^{4,2}$. These are the same 4-component spinors that we introduced for $\bbR^{4,1}$ in section \ref{sec:prelim_bulk:spinors}; however, they are now Weyl spinors, with different handedness for the lower-index and upper-index representations. We will identify the lower-index and upper-index spinors as left-handed and right-handed, respectively. The 4+2d gamma matrices couple the two representations, and are thus composed of lower-index and upper-index pieces. We can represent them in terms of the 4+1d matrices \eqref{eq:gamma} as:
\begin{align}
 \gamma^A_{ab} = \left(-iI_{ab}, \gamma^\mu_{ab} \right) \quad ; \quad \gamma_A^{ab} = \left(-iI^{ab}, \gamma_\mu^{ab} \right) \ .
\end{align}
The $\gamma^A_{ab}$ and $\gamma_A^{ab}$ span the spaces of lower-index and upper-index bispinors. Using the representation \eqref{eq:X} for a point $x\in dS_4$, we can now write the chiral projectors \eqref{eq:projectors} as:
\begin{align}
 P^L_{ab}(x) = -\frac{i}{2}X_A\gamma^A_{ab} \equiv -\frac{i}{2}X_{ab} \quad ; \quad 
 P_R^{ab}(x) = \frac{i}{2}X^A\gamma_A^{ab} \equiv \frac{i}{2}X^{ab} \ .
\end{align}
The left-handed field strength solutions \eqref{eq:spinor_solution} can now be written in 4+2d language as:
\begin{align}
 \varphi_{a_1 a_2\dots a_{2s}}(X) = \frac{M^L_{a_1}(X)M^L_{a_2}(X)\dots M^L_{a_{2s}}(X)}{(X\cdot\Xi)^{2s+1}} \ ; \quad
 M^L_a(X) \equiv -\frac{i}{2}\,X_{ab}M^b \ , \label{eq:spinor_solution_X}
\end{align}
and similarly for the right-handed solutions. The expression \eqref{eq:spinor_solution_X} is manifestly $SO(4,2)$-covariant with conformal weight $1$, and regular on the asymptotic section \eqref{eq:L}. The conformal weight $1-s$ of $\varphi_{\mu_1\nu_1\dots\mu_s\nu_s}$ is recovered in the translation to 3+1d tensors.

We have thus demonstrated that a spanning set of free-field solutions in $dS_4$ is regular through $\scri^\pm$ on the conformal compactification. This concludes the proof of Theorems \ref{thm:free_scalar}-\ref{thm:free_gauge} on the relation between antipodal symmetry and asymptotics. Note that the solutions \eqref{eq:scalar_solution} and \eqref{eq:spinor_solution} are not themselves antipodally symmetric. However, they can be combined with their antipodal images to form spanning sets of antipodally even/odd solutions. These will be closely related to the boundary-to-bulk propagators of section \ref{sec:propagators}.

\section{Interpretation in terms of $dS_4/\bbZ_2$} \label{sec:elliptic}

As discussed in the Introduction, antipodally-identified de Sitter space $dS_4/\bbZ_2$ is the quotient of $dS_4$ under the antipodal map $x\leftrightarrow -x$. One expects that antipodally symmetric fields in $dS_4$ can be interpreted as fields in $dS_4/\bbZ_2$. Let us flesh out the precise form of this statement. We will then use it to formulate Theorems  \ref{thm:free_scalar}-\ref{thm:free_gauge} in terms of 2-point functions on $dS_4/\bbZ_2$.

As a manifold, $dS_4/\bbZ_2$ is doubly-connected: it has incontractible cycles that correspond to paths between antipodal points in $dS_4$. It is also non-orientable, i.e. there is no global choice for the sign of $\epsilon^{\mu\nu\rho\sigma}$: since $\epsilon^{\mu\nu\rho\sigma}$ is antipodally odd in $dS_4$, it flips sign as one travels around an incontractible cycle in $dS_4/\bbZ_2$. Furthermore, since past and future have been identified, the metric of $dS_4/\bbZ_2$ does not admit a global time orientation. 

There are two ways to construct tensor fields on $dS_4/\bbZ_2$. Formally speaking, the fields can take values in two different line bundles over $dS_4/\bbZ_2$, which we will call the even bundle and the odd bundle. The even bundle is the trivial bundle of real/complex numbers at each point. The odd bundle is topologically non-trivial, such that the fiber (with all the field values in it) flips sign upon traversing an incontractible cycle. Clearly, antipodally even/odd fields on $dS_4$ correspond to even/odd fields on $dS_4/\bbZ_2$. In particular, the $dS_4/\bbZ_2$ metric is an even field, while the Levi-Civita tensor is an odd one. 

To be well-defined, interacting field equations in $dS_4/\bbZ_2$ must be such that powers of odd dynamical fields go together with powers of $\epsilon^{\mu\nu\rho\sigma}$. Thus, $dS_4/\bbZ_2$ only supports field theories that conserve P (and therefore CT), where the even/odd fields have even/odd intrinsic parity. Thus, solutions in $dS_4/\bbZ_2$ correspond to $dS_4$ solutions where the parity-even (parity-odd) fields are antipodally even (antipodally odd). The restriction to CT-preserving theories is not surprising: recall from section \ref{sec:prelim_bulk:geometry} that the antipodal map in $dS_4$ is an operation of the CT type. 

The conformal boundary of $dS_4/\bbZ_2$ is a single 3-sphere $\scrid$, resulting from the antipodal identification of $\scri^-$ and $\scri^+$. While $\scrid$ is of course orientable, it does not inherit a preferred orientation from the bulk. In particular, we've seen in section \ref{sec:prelim_scri:geometry} that $\scri^\pm$ are oppositely oriented (unless one chooses a preferred time direction, which is impossible in the $dS_4/\bbZ_2$ context). As with the bulk, one can define even and odd fields intrinsically on $\scrid$, where the even/odd distinction refers both to the fields' intrinsic parity and to their antipodal symmetry on $\scri^\pm$. There are incontractible cycles in $dS_4/\bbZ_2$ that connect a point on $\scrid$ to itself, via the bulk. Odd fields on $\scrid$ flip their sign upon traversing such a cycle.

Having understood the geometry of $dS_4/\bbZ_2$ and its boundary $\scrid$, we can reformulate the statements of Theorems \ref{thm:free_scalar}-\ref{thm:free_gauge} as follows:
\begin{enumerate}
 \item An even/odd scalar field on $dS_4/\bbZ_2$ that satisfies the field equation $(\Box - 2)\varphi = 0$ has vanishing Dirichlet/Neumann boundary data $\phi$/$\pi$ (and is determined by the boundary data of the other type).
 \item An even/odd spin-$s$ gauge field on $dS_4/\bbZ_2$ that satisfies the field equations \eqref{eq:Maxwell_diff}-\eqref{eq:Weyl_diff} has vanishing electric/magnetic boundary data $\calE_{\mu_1\dots\mu_s}$/$\calB_{\mu_1\dots\mu_s}$ (and is determined by the boundary data of the other type).
\end{enumerate}
This can be further reformulated in terms of 2-point functions on $\scrid$:
\begin{enumerate}
 \item An even/odd scalar field on $dS_4/\bbZ_2$ with field equation $(\Box - 2)\varphi = 0$ has a vanishing Neumann/Dirichlet 2-point function on $\scrid$, while the 2-point function of the other type is ill-defined.
 \item An even/odd spin-$s$ gauge field on $dS_4/\bbZ_2$ with field equations \eqref{eq:Maxwell_diff}-\eqref{eq:Weyl_diff} has a vanishing magnetic/electric 2-point function on $\scrid$, while the 2-point function of the other type is ill-defined.
\end{enumerate}
The result persists in interacting theories, since the 2-point functions are determined by the free equations. Generically, the interaction will produce finite $n$-point functions with $n>2$. However, in the special case of Vasiliev gravity, we will see in section \ref{sec:vasiliev} that the $n$-point functions have the same singular behavior for $n>2$ as they do for $n=2$.

\section{Initial value problem for interacting theories in $dS_4/\bbZ_2$} \label{sec:initial_value}

We will now use the free-field result of section \ref{sec:relations} to formulate a well-defined initial value problem for interacting gauge fields in $dS_4/\bbZ_2$. It is helpful to first formulate and prove the statement in terms of antipodally symmetric fields on $dS_4$. The $dS_4/\bbZ_2$ statement will be given at the end of the section. As we recall from section \ref{sec:elliptic}, only interactions that preserve P (and thus CT) respect the antipodal symmetry. For such theories, we find the following result in $dS_4$:
\begin{theorem}[Existence and uniqueness of perturbative solutions] \label{thm:interacting}
 Consider a classical parity-invariant field theory in 3+1 dimensions, which admits a perturbative expansion around empty de Sitter space $dS_4$. Assume that all linear field perturbations are $m^2=2$ scalars and gauge fields, i.e. that they satisfy the field equation \eqref{eq:h_diff} for the appropriate spin. Assume further that all fields have a definite intrinsic parity $\pm 1$. Fix on $\scri^\pm$ an antipodally even (antipodally odd) configuration of $\pi$/$\calB_{\mu_1\dots\mu_s}$ ($\phi$/$\calE_{\mu_1\dots\mu_s}$) boundary data for the parity-even (parity-odd) fields. Then, at each order of perturbation theory, there exists a unique bulk solution with the same antipodal symmetry as the boundary data (other solutions lacking the symmetry may also exist).
\end{theorem}
The theorem covers a wide range of theories, including: 
\begin{itemize}
 \item Theories of $m^2=2$ scalars and Maxwell/Yang-Mills fields on a fixed $dS_4$ metric.
 \item General Relativity with $\Lambda>0$, coupled to any of the above matter fields.
 \item Parity-invariant Vasiliev theories of interacting higher-spin gauge fields with $\Lambda>0$.
\end{itemize}
Note that the same field can sometimes be taken as either parity-even or parity-odd (though one must make a choice when going over to $dS_4/\bbZ_2$!). This is the case for scalar fields with an even potential in standard $s\leq 2$ theories, as well as for Maxwell fields.

We now turn to prove the theorem. In the interest of readability, we only present the proof for the case where all fields are parity-even. The field equations can then be written with no $\epsilon^{\mu\nu\rho\sigma}$ factors, and the boundary data prescribed in the theorem is $\{\pi,\calB_{\mu_1\dots\mu_s}\}$. The proof with parity-odd fields is analogous.
\begin{proof}[Proof of part (a)]
 We prove the statement by induction, order by order in perturbation theory. 

 At first order, consider the solution with boundary data $\{\phi=0,\pi,\calB_{\mu_1\dots\mu_s},\calE_{\mu_1\dots\mu_s}=0\}$ on $\scri^-$. This solution must exist by Cauchy evolution. By Theorems \ref{thm:free_scalar}-\ref{thm:free_gauge}, it is antipodally even, and satisfies the same boundary conditions (in particular, the required ones on $\{\pi,\calB_{\mu_1\dots\mu_s}\}$) at $\scri^+$. There are also other solutions that satisfy the given boundary conditions on $\{\pi,\calB_{\mu_1\dots\mu_s}\}$, with arbitrary antipodally \emph{odd} boundary values of $\{\phi,\calE_{\mu_1\dots\mu_s}\}$. However, the antipodally even solution - the one with vanishing $\{\phi,\calE_{\mu_1\dots\mu_s}\}$ - is unique.

 Now, assume that the statement holds for the first $n-1$ orders in perturbation theory. Let us fix the solution at these orders to the unique antipodally even one. The field equations for the $n$'th-order fields are just the linear equations \eqref{eq:h_diff}, but with source terms on the right-hand side. These source terms must be constructed covariantly out of the lower-order fields, the background metric $g_{\mu\nu}$ and the background covariant derivative $\nabla_\mu$. These objects are all antipodally even; therefore, the source terms constructed from them are also antipodally even. Now, consider an arbitrary solution to the $n$'th-order equations, e.g. the one with vanishing $\{\phi,\pi,\calB_{\mu_1\dots\mu_s},\calE_{\mu_1\dots\mu_s}\}$ on $\scri^-$. The antipodal image of this is also a solution, due to the symmetry of the source terms. Since the equations are linear, we can take the average of the two antipodal images, producing an antipodally even solution (which does not yet satisfy the required boundary conditions). The \emph{general} $n$'th-order solution can be obtained from this by adding a solution to the \emph{free} field equations. Now, by Theorems \ref{thm:free_scalar}-\ref{thm:free_gauge}, the free antipodally even solutions are in one-to-one correspondence with $\{\pi,\calB_{\mu_1\dots\mu_s}\}$ boundary data. Thus, we can add a unique free even solution that will fix $\{\pi,\calB_{\mu_1\dots\mu_s}\}$ to the required values. The remaining freedom is to add an antipodally \emph{odd} free solution, which by Theorems \ref{thm:free_scalar}-\ref{thm:free_gauge} will have vanishing $\{\pi,\calB_{\mu_1\dots\mu_s}\}$. However, such an addition would spoil the antipodal symmetry. We conclude that the antipodally even solution with given $\{\pi,\calB_{\mu_1\dots\mu_s}\}$ values is unique.
\end{proof}

In $dS_4/\bbZ_2$, Theorem \ref{thm:interacting} becomes the statement of an initial value problem:
\begin{itemize}
 \item Consider a field theory as in Theorem \ref{thm:interacting}. Fix a configuration of $\pi$/$\calB_{\mu_1\dots\mu_s}$ ($\phi$/$\calE_{\mu_1\dots\mu_s}$) boundary data on $\scrid$ for the parity-even (parity-odd) fields. Then there exists a unique bulk solution in $dS_4/\bbZ_2$. 
\end{itemize}

\section{Boundary-to-bulk propagators in $dS_4/\bbZ_2$} \label{sec:propagators}

In this section, we present boundary-to-bulk propagators for gauge fields in $dS_4/\bbZ_2$. In the scalar case, we present both the antipodally even (i.e. Neumann) and the antipodally odd (i.e. Dirichlet) propagators. For gauge fields with spin $s>0$, we present only the antipodally even (i.e. magnetic) propagators. We will use these propagators in our treatment of Vasiliev gravity in section \ref{sec:vasiliev}.

\subsection{Scalar propagators}

The boundary-to-bulk propagators for scalars in $EAdS_4$ are well-known \cite{Witten:1998qj,Freedman:1998tz}. For our $m^2=2$ case, they read, in the 4+1d language:
\begin{align}
 \text{Neumann:} \quad \varphi^E(x;\ell) \sim \frac{1}{x\cdot\ell} \quad ; \quad \text{Dirichlet:} \quad \tilde\varphi^E(x;\ell) \sim \frac{1}{(x\cdot\ell)^2} \ . \label{eq:euclidean_scalar}
\end{align}
Here, $x^\mu$ is a unit timelike vector in $\bbR^{4,1}$ representing the bulk point in $EAdS_4$, while $\ell^\mu$ is a null vector representing the boundary point. The propagators \eqref{eq:euclidean_scalar} are solutions to the free-field equation \eqref{eq:scalar_diff}. For the Neumann/Dirichlet propagator, the $\pi$/$\phi$ boundary data is a delta-function at the point encoded by $\ell^\mu$, while the other type of boundary data is fixed by regularity on $EAdS_4$. 

When translating the propagators \eqref{eq:euclidean_scalar} into $dS_4$, the vector $x^\mu$ representing the bulk point becomes spacelike. The denominators in \eqref{eq:euclidean_scalar} now vanish on the lightcone of the boundary point, necessitating an $i\varepsilon$ prescription. For future-pointing $\ell^\mu$, the two prescriptions $x\cdot\ell \rightarrow x\cdot\ell \pm i\varepsilon$ yield the positive-frequency and negative-frequency propagators in the Bunch-Davies sense. These two prescriptions are antipodal images of each other. We can therefore obtain antipodally symmetric propagators by superposing them. In accordance with Theorem \ref{thm:free_scalar}, the antipodally even/odd propagators should have a delta-function for the Neumann/Dirichlet boundary data, with the other type of boundary data \emph{vanishing}. Since the field only propagates along lightrays, this implies that the propagators must vanish away from the lightcone $x\cdot\ell = 0$ of the boundary point. Thus, the propagators in $dS_4/\bbZ_2$ take the form:
\begin{align}
 \text{Even/Neumann:} \quad &\varphi(x;\ell) = \frac{1}{4\pi}\delta(x\cdot\ell) 
   = -\frac{1}{8\pi^2 i}\left(\frac{1}{x\cdot\ell + i\varepsilon} - \frac{1}{x\cdot\ell - i\varepsilon}\right) \ ; \label{eq:scalar_propagator_even} \\
 \text{Odd/Dirichlet:} \quad &\tilde\varphi(x;\ell) = \frac{1}{4\pi}\delta'(x\cdot\ell) 
   = \frac{1}{8\pi^2 i}\left(\frac{1}{(x\cdot\ell + i\varepsilon)^2} - \frac{1}{(x\cdot\ell - i\varepsilon)^2}\right) \ . \label{eq:scalar_propagator_odd}
\end{align}
Note that taking $x$ to the boundary in the propagators \eqref{eq:scalar_propagator_even}-\eqref{eq:scalar_propagator_odd} yields vanishing 2-point functions, as required by Theorem \ref{thm:free_scalar} and its reformulation in section \ref{sec:elliptic}.

The normalizations in \eqref{eq:scalar_propagator_even}-\eqref{eq:scalar_propagator_odd} can be verified by integrating the propagators over the boundary. To do this, choose a frame in $\bbR^{4,1}$ such that $\ell^\mu$ takes the form:
\begin{align}
 \ell^\mu = e_0^\mu + e_4^\mu \ . \label{eq:ell}
\end{align}
The bulk point $x^\mu$ can be parametrized as:
\begin{align}
 x^\mu = \sinh\eta\,e_0^\mu + \cosh\eta\left(\cos\chi\,e_4^\mu + \sin\chi\left(\cos\theta\,e_3^\mu + \sin\theta(\cos\phi\,e_1^\mu + \sin\phi\,e_2^\mu)\right)\right) \ .
 \label{eq:x}
\end{align}
As a hypersurface approaching e.g. $\scri^+$, we choose a constant-$\eta$ slice with $\eta\rightarrow\infty$. This is a 3-sphere with radius $\cosh\eta$. It becomes a unit 3-sphere upon rescaling with e.g. $z=1/\sinh\eta$ at $z\rightarrow 0$ (this choice of $z$ has the correct signs at $\scri^\pm$). The scalar product in the delta functions \eqref{eq:scalar_propagator_even}-\eqref{eq:scalar_propagator_odd} reads:
\begin{align}
 u \equiv x\cdot\ell = \cosh\eta\cos\chi - \sinh\eta \ . \label{eq:u}
\end{align}
Since this has no $(\theta,\phi)$ dependence, we can write the volume element on the 3-sphere as:
\begin{align}
 dV = 4\pi\sin^2\chi\, d\chi = -\frac{4\pi\sqrt{1 - u^2 - 2u\sinh\eta}}{\cosh^2\eta}\,du \longrightarrow -4\pi z^2 du\sqrt{1 - u^2 - 2u/z} \ , \label{eq:dV}
\end{align}
where the arrow denotes the $z\rightarrow 0$ limit. Integrating the delta functions in \eqref{eq:scalar_propagator_even}-\eqref{eq:scalar_propagator_odd} on the $(\chi,\theta,\phi)$ 3-sphere, we get:
\begin{align}
 \int dV \delta(x\cdot\ell) &= 4\pi z^2\left.\sqrt{1 - u^2 - 2u/z}\right|_{u=0} = 4\pi z^2 \ ; \\ 
 \int dV \delta'(x\cdot\ell) &= -4\pi z^2\left.\frac{d}{du}\sqrt{1 - u^2 - 2u/z}\right|_{u=0} = 4\pi z \label{eq:int_delta_prime} \ ,
\end{align}
where in \eqref{eq:int_delta_prime} we kept only the leading term in $z$. This shows that the propagators \eqref{eq:scalar_propagator_even}-\eqref{eq:scalar_propagator_odd} yield normalized delta functions for the $\pi$/$\phi$ boundary data, respectively.

\subsection{Gauge potential propagators}

We now turn to the magnetic propagators for the spin-$s$ gauge potentials. In addition to the boundary point encoded by $\ell^\mu$, the propagator must now depend on a (symmetric, traceless) polarization tensor on $\scri$. Without loss of generality, we can take this to have the form $\lambda^{\mu_1}\dots\lambda^{\mu_s}$, where $\lambda^\mu$ is a (complex) null vector on $\scri$. In the 4+1d picture, this means that $\lambda^\mu$ is a null vector orthogonal to $\ell^\mu$, defined up to multiples of $\ell^\mu$. This data can be neatly encoded in a totally null \emph{bivector} $M^{\mu\nu} = \ell^\mu\wedge\lambda^\nu$, which has the properties:
\begin{align}
 M^{\mu\nu} = -M^{\nu\mu}\ , \quad M^{\mu\nu}M_{\nu\rho} = 0\ , \quad M^{\mu\nu}\ell_\nu = 0 \ .
\end{align}
In $EAdS_4$, the propagators have been worked out in \cite{Mikhailov:2002bp}. In our language, they read:
\begin{align}
 h^E_{\mu_1\dots\mu_s}(x;\ell,M) \sim \frac{M_{\mu_1\nu_1}x^{\nu_1}\dots M_{\mu_s\nu_s}x^{\nu_s}}{(x\cdot\ell)^{2s+1}} \ . \label{eq:gauge_euclidean}
\end{align}
These are solutions to the free field equations \eqref{eq:h_diff_simple} in transverse traceless gauge. Note that $M^{\mu\nu}x_\nu$ is automatically in the tangent space of $dS_4$, i.e. orthogonal to $x^\mu$. 

As in the scalar case, when translating the propagators \eqref{eq:gauge_euclidean} to $dS_4/\bbZ_2$, we must choose the delta-function-like combination of the two $x\cdot\ell \rightarrow x\cdot\ell \pm i\varepsilon$ prescriptions. This gives:
\begin{align}
 \begin{split}
   &h_{\mu_1\dots\mu_s}(x;\ell,M) = \frac{1}{C(s)}\, \delta^{(2s)}(x\cdot\ell)M_{\mu_1\nu_1}x^{\nu_1}\dots M_{\mu_s\nu_s}x^{\nu_s} \\
   &\quad = -\frac{(2s)!}{2\pi i C(s)}
      \left(\frac{1}{(x\cdot\ell + i\varepsilon)^{2s+1}} - \frac{1}{(x\cdot\ell - i\varepsilon)^{2s+1}} \right) M_{\mu_1\nu_1}x^{\nu_1}\dots M_{\mu_s\nu_s}x^{\nu_s} \ ,
 \end{split} \label{eq:gauge_propagator}
\end{align}
where $C(s)$ is a normalization factor, and $\delta^{(2s)}$ is the $2s$-th derivative of the delta function. The propagator \eqref{eq:gauge_propagator} is antipodally even, and therefore has purely-magnetic boundary data. 

At the boundary, the tangential components of the propagator \eqref{eq:gauge_propagator} scale as $z^{2-2s}$ (in a coordinate basis on $\scri$); the coefficient is $\lambda_{\mu_1}\dots\lambda_{\mu_s}$ times a delta function at the point encoded by $\ell^\mu$. In a previous version of this manuscript, a wrong value was given for the normalization coefficient $C(s)$ that leads to a normalized delta function. A more careful analysis has been carried out in \cite{Halpern:2015zia}, yielding the value:
\begin{align}
 C(s) = 2\pi(-1)^{s+1}\frac{(2s)!(2s-3)!!}{s!} \ ,
\end{align}
where the double factorial $(2s-3)!!$ is defined as $1$ for $s=1$ and $1\cdot 3\cdot 5\cdot\dots\cdot (2s-3)$ for $s\geq 2$.

\subsection{Gauge potential and field strength propagators in spinor form}

To derive the field strengths from the gauge potential propagators \eqref{eq:gauge_propagator}, it is helpful to first rewrite them in spinor form. The polarization bivector $M^{\mu\nu}$ gets translated into a 4+1d spinor $M^a$ via:
\begin{align}
 M^{\mu\nu} = \frac{1}{4}\gamma^{\mu\nu}_{ab}M^a M^b \ .
\end{align}
One can then show that the vector $M^{\mu\nu}x_\nu$ becomes:
\begin{align}
 M^{\mu\nu}x_\nu = -\frac{i}{2}\,\gamma^\mu_{\alpha\dot\alpha}M_L^\alpha(x)M_R^{\dot\alpha}(x) \ ,
\end{align}
where we recall that $M_L^\alpha(x)$ and $M_R^{\dot\alpha}(x)$ are the projections of $M^a$ onto the left-handed and right-handed spinor spaces at $x$. The gauge-potential propagator \eqref{eq:gauge_propagator} then becomes:
\begin{align}
 h_{\alpha_1\dots\alpha_s\dot\alpha_1\dots\dot\alpha_s}(x;\ell,M) = \frac{i^s}{C(s)}\, \delta^{(2s)}(x\cdot\ell)
     M^L_{\alpha_1}(x)\dots M^L_{\alpha_s}(x)M^R_{\dot\alpha_1}(x)\dots M^R_{\dot\alpha_s}(x) \ . \label{eq:gauge_propagator_spinor}
\end{align}
The left-handed and right-handed field strength propagators can now be derived as in \eqref{eq:phi_h_spinor} to give:
\begin{align}
 \begin{split}
   &\varphi_{\alpha_1\dots\alpha_{2s}}(x;\ell,M) = \frac{(-1)^s (2s)!}{2^{s+1}\, s!\, C(s)}\, \delta^{(2s)}(x\cdot\ell) M^L_{\alpha_1}(x)\dots M^L_{\alpha_{2s}}(x) \\
   &\quad = \frac{(-1)^{s+1} \left((2s)!\right)^2}{2\pi i\cdot 2^{s+1}\, s!\, C(s)} 
     \left(\frac{1}{(x\cdot\ell + i\varepsilon)^{2s+1}} - \frac{1}{(x\cdot\ell - i\varepsilon)^{2s+1}} \right) M^L_{\alpha_1}(x)\dots M^L_{\alpha_{2s}}(x) \ ; \\
   &\varphi_{\dot\alpha_1\dots\dot\alpha_{2s}}(x;\ell,M) = \frac{(-1)^s (2s)!}{2^{s+1}\, s!\, C(s)}\, \delta^{(2s)}(x\cdot\ell) M^R_{\dot\alpha_1}(x)\dots M^R_{\dot\alpha_{2s}}(x) \\
   &\quad = \frac{(-1)^{s+1} \left((2s)!\right)^2}{2\pi i\cdot 2^{s+1}\, s!\, C(s)} 
     \left(\frac{1}{(x\cdot\ell + i\varepsilon)^{2s+1}} - \frac{1}{(x\cdot\ell - i\varepsilon)^{2s+1}} \right) M^R_{\dot\alpha_1}(x)\dots M^R_{\dot\alpha_{2s}}(x) \ ,
 \end{split} \label{eq:field_strength_propagator}
\end{align}
where the individual terms on the second lines are the propagators with positive/negative frequency in the Bunch-Davies sense. The derivation of \eqref{eq:field_strength_propagator} from \eqref{eq:gauge_propagator_spinor} follows from the relations:
\begin{align}
 \begin{split}
   M_R^{\dot\alpha}(x)\nabla_{\alpha\dot\alpha}(x\cdot\ell) &= -i(x\cdot\ell) M^L_\alpha(x) \ ; \\ 
   \nabla_{\alpha\dot\alpha} M_R^{\dot\alpha}(x) &= -2iM^L_\alpha(x) \ ; \\
   M_R^{\dot\alpha}(x)\nabla_{\alpha\dot\alpha} M_R^{\dot\beta}(x) &= -iM^L_\alpha(x) M_R^{\dot\beta}(x) \ ; \\
   M_R^{\dot\alpha}(x)\nabla_{\alpha\dot\alpha} M_L^\beta(x) &= 0 \ ,
 \end{split} \label{eq:diff_M_R_spinor}
\end{align} 
along with their counterparts of opposite chirality.

\section{Higher-spin gravity in $dS_4/\bbZ_2$} \label{sec:vasiliev}

\subsection{Choice of theory and boundary conditions}

We now turn to discuss Vasiliev's higher-spin gravity in $dS_4/\bbZ_2$. The theory comes in a variety of versions. In this paper, we focus on purely bosonic ones. As discussed in section \ref{sec:elliptic}, the $dS_4/\bbZ_2$ context further restricts us to parity-invariant theories. This leaves us with just four possibilities, distinguished by two binary choices. The first choice is between a minimal theory (even spins only) and a non-minimal one (both even and odd spins). We will treat these two options simultaneously, with the minimal $n$-point functions forming a subset of the non-minimal ones. The second choice is between type-A (parity-even scalar field) and type-B (parity-odd scalar field); the $s>0$ gauge fields are always parity-even. The dS/CFT model of \cite{Anninos:2011ui} uses the minimal type-A theory. Here, we consider all four of the parity-invariant bosonic versions.

Having chosen the bulk theory, one can work with different choices of boundary conditions. In ordinary (A)dS, the possibilities are as follows \cite{Giombi:2013yva}. For the scalar field, we can use either Dirichlet or Neumann boundary conditions, i.e. we can fix either $\phi(x)$ or $\pi(x)$ on $\scrid$. For the $s>0$ fields, we can fix any linear combination of $\calE_{\mu_1\dots\mu_s}(x)$ and $\calB_{\mu_1\dots\mu_s}(x)$. Magnetic boundary conditions (fixing $\calB_{\mu_1\dots\mu_s}$) and electric ones (fixing $\calE_{\mu_1\dots\mu_s}$) are the two limiting cases. 

Normally, these different boundary conditions are just different parametrizations of the same bulk solutions (or amplitudes). In particular, the Dirichlet/Neumann or magnetic/electric conditions are related to each other by Legendre transforms. In $dS_4/\bbZ_2$, however, these transforms become singular. Indeed, as we've seen from Theorems \ref{thm:free_scalar}-\ref{thm:free_gauge}, the 2-point functions for a particular choice of boundary conditions vanish. According to the fields' intrinsic parities, these are magnetic conditions on the gauge fields and Neumann/Dirichlet conditions on the scalar field, in the type-A/type-B theory respectively. Incidentally, these are precisely the boundary conditions that preserve the higher-spin symmetry \cite{Vasiliev:2012vf}, and that correspond to free boundary theories in AdS/CFT \cite{Giombi:2013yva}. 

In the following, we will focus on this particular choice of boundary conditions, and argue that not only the 2-point functions, but all the $n$-point functions vanish. This means that the opposite types of boundary data (electric for the gauge fields, Dirichlet/Neumann for the type-A/type-B scalar) continue to vanish at all orders in perturbation theory. As a consequence, the $n$-point functions with these data as boundary conditions are ill-defined. The same conclusion applies to mixed boundary conditions that fix a combination of $\calE_{\mu_1\dots\mu_s}$ and $\calB_{\mu_1\dots\mu_s}$.

\subsection{Higher-spin framework}

In Vasiliev gravity, one augments spacetime with an internal twistor space. In standard treatments, this twistor space is rigidly decomposed into left-handed and right-handed spinor spaces. Such a formalism is well-suited for calculations in Poincare coordinates, but it is not covariant under the full $SO(4,1)$ de Sitter group. Moreover, it cannot be used in $dS_4/\bbZ_2$, since the latter is non-orientable. A simple alternative is to identify the internal twistor space with the global space of $SO(4,1)$ spinors from section \ref{sec:prelim_bulk:spinors}. The price is that the decomposition into left-handed and right-handed spinors is now $x$-dependent, governed by the $P_{L/R}(x)$ projectors from \eqref{eq:projectors} (and is only possible locally in $dS_4/\bbZ_2$, since the $P_{L/R}$ are interchanged by the antipodal map). A formalism for such generalized gauges is given in \cite{Vasiliev:2001wa,Didenko:2012vh}. It involves a ``compensator field'' which reduces the symmetry of the internal space from $SO(4,1)$ to $SO(3,1)$. In our case, this role is played by the radius-vector $x^\mu$.

The detailed framework is as follows. The higher-spin algebra is generated by twistor variables $Y^a$, subject to the star product:
\begin{align}
 Y_a\star Y_b = Y_a Y_b + iI_{ab} \ .
\end{align}
The $SO(4,1)$ generators and their commutators are given by:
\begin{align}
 T_{\mu\nu} = \frac{i}{8}\gamma_{\mu\nu}^{ab}Y_a Y_b \quad ; \quad [T^{\mu\nu}, T_{\rho\sigma}]_\star = 4\delta^{[\mu}_{[\rho} T^{\nu]}{}_{\sigma]} \ .
\end{align}
As discussed above, we use a gauge where the space of the $Y^a$ is identified with the global $SO(4,1)$ spinor space in pure $dS_4$ (or $dS_4/\bbZ_2$). In this gauge, the background higher-spin connection $\Omega$ vanishes. Instead, we have the frame one-form $\Sigma(x) = T_{\mu\nu}dx^\mu x^\nu$, which encodes the translation generators at the point $x$. 

Our treatment of perturbations around $dS_4/\bbZ_2$ will focus on the zero-form master field $B(x,Y)$, which encodes the field strengths for all spins along with their spacetime derivatives. It will suffice to work with $B(x,Y)$ at the linearized level. In our $\Omega = 0$ gauge, the free field equation for $B$ reads simply:
\begin{align}
 dB - 2B\star\Sigma = 0 \quad ; \quad \Sigma = T_{\mu\nu}dx^\mu x^\nu = \frac{i}{8}dx^\mu x^\nu\gamma_{\mu\nu}^{ab}Y_a Y_b \ . \label{eq:master_diff}
\end{align}
The master field $B(x,Y)$ is parity-even in the type-A theory, and parity-odd in the type-B theory. This is despite the fact that the $s>0$ component gauge fields are parity-even in both cases: in the type-B case, there is a handedness-dependent sign factor in the translation between the master field and the component fields. In $dS_4/\bbZ_2$, the intrinsic parities are translated into antipodal symmetry signs: $B(x,Y)$ is antipodally even/odd in the type-A/type-B theory, even though the $s>0$ component fields are always antipodally even.

\subsection{Master-field propagators}

The scalar propagators \eqref{eq:scalar_propagator_even}-\eqref{eq:scalar_propagator_odd} and the field strength propagators \eqref{eq:field_strength_propagator} can be embedded (up to normalizations) into a pair of master fields that satisfy eq. \eqref{eq:master_diff}:
\begin{align}
 \begin{split}
   B(x,Y;\ell,M) \sim{}& \frac{1}{x\cdot\ell + i\varepsilon}\, \exp\frac{\gamma^{ab}_{\mu\nu}\ell^\mu x^\nu Y_a Y_b}{2i(x\cdot\ell + i\varepsilon)} 
    \left(\exp\frac{iP_L^{ab}(x)M_a Y_b}{x\cdot\ell + i\varepsilon} \pm \exp\frac{iP_R^{ab}(x)M_a Y_b}{x\cdot\ell + i\varepsilon} \right) \\
    &+ (Y^a \longrightarrow -Y^a) - (i\varepsilon \longrightarrow -i\varepsilon) \ ; 
 \end{split} \label{eq:master_propagator} \\
 \tilde B(x,Y;\ell) \sim{}& \frac{1}{(x\cdot\ell + i\varepsilon)^2}\left(1 + \frac{\gamma^{ab}_{\mu\nu}\ell^\mu x^\nu Y_a Y_b}{2i(x\cdot\ell + i\varepsilon)} \right)
   \exp\frac{\gamma^{ab}_{\mu\nu}\ell^\mu x^\nu Y_a Y_b}{2i(x\cdot\ell + i\varepsilon)} - (i\varepsilon \longrightarrow -i\varepsilon) \ . \label{eq:master_propagator_tilde}
\end{align}
These master-field propagators are generating functions in $M^a$ and $Y^a$. The different powers of $M^a$ encode the boundary data for the corresponding spins. The different powers and handedness components of $Y^a$ encode the field strengths of different spins and their spacetime derivatives; specifically, the spacetime derivatives are associated with the factors of $\gamma^{ab}_{\mu\nu}\ell^\mu x^\nu Y_a Y_b$. The $Y^a\rightarrow -Y^a$ symmetrization picks out the integer spins.

In the type-A theory, we choose the $+$ sign in the propagator \eqref{eq:master_propagator}. It is then antipodally even, and contains the Neumann scalar propagator \eqref{eq:scalar_propagator_even} and the magnetic gauge field propagators \eqref{eq:field_strength_propagator}. In the type-B theory, we choose the $-$ sign in the propagator \eqref{eq:master_propagator}. It is then antipodally odd, and contains \emph{only} the magnetic gauge field propagators \eqref{eq:field_strength_propagator}; the Dirichlet scalar propagator \eqref{eq:scalar_propagator_odd} is encoded separately in the antipodally odd master field \eqref{eq:master_propagator_tilde}.

Without the $i\varepsilon \rightarrow -i\varepsilon$ antisymmetrization, \eqref{eq:master_propagator}-\eqref{eq:master_propagator_tilde} are just the propagators from \cite{Giombi:2009wh,Giombi:2010vg,Didenko:2012tv,Didenko:2013bj} rewritten in our covariant gauge, with positive frequency in the Bunch-Davies sense. As in section \ref{sec:propagators}, the $i\varepsilon \rightarrow -i\varepsilon$ antisymmetrization imposes antipodal symmetry, while turning the component fields into distributions with support on the $x\cdot\ell = 0$ lightcone.

\subsection{$n$-point functions}

We can now plug the master-field propagators \eqref{eq:master_propagator}-\eqref{eq:master_propagator_tilde} into the $n$-point function calculations of \cite{Giombi:2010vg,Didenko:2012tv,Didenko:2013bj}. In \cite{Giombi:2010vg}, the 3-point function is calculated in a gauge where the higher-spin connection \emph{as well as} the frame field $\Sigma$ vanish. In this gauge, the $B$ master-field propagators become $x$-independent; the propagators' value at all $x$ is given by their value at an arbitrary base point $x_0$ in the original, ``physical'' gauge \eqref{eq:master_propagator}-\eqref{eq:master_propagator_tilde}. The 3-point function is then calculated as a bilinear functional of these propagators. Crucially, this means that the 3-point function is expressed as a functional of the physical-gauge propagators \eqref{eq:master_propagator}-\eqref{eq:master_propagator_tilde} at a single, arbitrary point $x_0$: there is no need to integrate over the location of the ``interaction point'' in spacetime.

Now, recall that the $dS_4/\bbZ_2$ propagators \eqref{eq:master_propagator}-\eqref{eq:master_propagator_tilde} \emph{vanish} at a generic point (i.e. a point that isn't on the lightcone of the boundary source). Therefore, in $dS_4/\bbZ_2$, the calculation of \cite{Giombi:2010vg} with a generic base point $x_0$ for the gauge transformation will yield a vanishing 3-point function! The argument must be made with some care, due to the singular distributional nature of the propagators \eqref{eq:master_propagator}-\eqref{eq:master_propagator_tilde}. Naively, if the propagators vanish at one point, then one can use the field equation \eqref{eq:master_diff} to show that they vanish everywhere. Thus, the propagators should really be defined as the limit of a sequence of non-singular fields, which do not vanish anywhere. However, the conclusion remains intact: in the limit, the propagators away from the lightcone become arbitrarily small, and one still gets zero when plugging them into the 3-point function calculation of \cite{Giombi:2010vg}.

Similarly, in the $n$-point function calculations of \cite{Didenko:2012tv,Didenko:2013bj}, the result is obtained as a multilinear functional of the $B$ master-field propagators \emph{at an arbitrary point}. Since our $dS_4/\bbZ_2$ propagators vanish away from the boundary sources' lightcones, we conclude that all the $n$-point functions vanish.

\section{Discussion} \label{sec:discuss}

In this paper, we studied the relations between asymptotic boundary data, parity and antipodal symmetry for gauge fields in $dS_4$. We constructed a perturbatively well-posed initial value problem at the conformal boundary of elliptic de Sitter space $dS_4/\bbZ_2$. The results apply to realistic theories such as Yang-Mills and General Relativity, as well as to Vasiliev's higher-spin gravity. The latter features as the bulk theory in a family of AdS/CFT dualities, which appear particularly suited for reformulation with a positive cosmological constant. We explored the possibility of a dS/CFT duality that calculates Lorentzian higher-spin ``transition amplitudes'' in $dS_4/\bbZ_2$. We found that this notion is empty, since the $n$-point functions are all either zero or ill-defined, depending on the choice of boundary data. The same is true for 2-point functions in any theory of free or interacting gauge fields. However, the conclusion for the higher $n$-point functions seems to be special to Vasiliev gravity. For instance, using the propagator \eqref{eq:scalar_propagator_even}, one can compute the 3-point function for an $m^2=2$ scalar with a simple $\varphi^3$ interaction, and the result is finite. 

Our proof for the vanishing of the higher-spin $n$-point functions is only as good as the $n$-point function calculations of \cite{Giombi:2010vg,Didenko:2012tv,Didenko:2013bj}. In the first of these references, only the 3-point function is computed. In the other two, one employs an indirect argument based on higher-spin symmetry, which is only conjectured to agree with the explicit solution of Vasiliev's field equations. When more complete calculations appear, it will be possible to test our result against them.

As discussed in the Introduction, elliptic de Sitter space remains a fascinating testing ground for ideas in quantum gravity, in particular horizon complementarity. There is much to understand about fields in this spacetime. For example, it appears that they cannot be quantized globally, but only after choosing an observer with his associated cosmological horizons. We pursue these issues in a separate work \cite{YashaLucas}. Our null result for the higher-spin correlators fits neatly into this picture, as another piece of evidence that one cannot do global physics in $dS_4/\bbZ_2$.

\section*{Acknowledgements}

I am grateful to Abhay Ashtekar, Beatrice Bonga, Lucas Hackl, Simone Giombi, Gim Seng Ng and Wolfgang Wieland for discussions. Research at Perimeter Institute is supported by the Government of Canada through Industry Canada and by the Province of Ontario through the Ministry of Research \& Innovation. YN also acknowledges support of funding from NSERC Discovery grants. The first version of the manuscript was produced at Penn State University, where it was supported in part by the NSF grant PHY-1205388 and the Eberly Research Funds of Penn State.

\end{document}